\documentclass[aps,prb,twocolumn,superscriptaddress,showpacs]{revtex4}

\usepackage{graphicx}% Include figure files
\usepackage{dcolumn}% Align table columns on decimal point
\usepackage{bm}% bold math

\begin{document}

\title{Low-temperature heat transport of Nd$_{2-x}$Ce$_x$CuO$_4$ single crystals}

\author{X. Zhao}
\email{xiazhao@ustc.edu.cn}

\affiliation{School of Physical Sciences, University of Science
and Technology of China, Hefei, Anhui 230026, People's Republic of
China}

\author{Z. Y. Zhao}
\email{zzhao20@utk.edu}

\affiliation{Hefei National Laboratory for Physical Sciences at
Microscale, University of Science and Technology of China, Hefei,
Anhui 230026, People's Republic of China}

\affiliation{Department of Physics and Astronomy, University of
Tennessee, Knoxville, Tennessee 37996-1200, USA}

\affiliation{Materials Science and Technology Division, Oak Ridge National Laboratory, Oak Ridge, Tennessee 37831, USA}

\author{B. Ni}
\affiliation{Hefei National Laboratory for Physical Sciences at
Microscale, University of Science and Technology of China, Hefei,
Anhui 230026, People's Republic of China}

\author{J. C. Wu}
\affiliation{Hefei National Laboratory for Physical Sciences at
Microscale, University of Science and Technology of China, Hefei,
Anhui 230026, People's Republic of China}

\author{F. B. Zhang}
\affiliation{Hefei National Laboratory for Physical Sciences at
Microscale, University of Science and Technology of China, Hefei,
Anhui 230026, People's Republic of China}

\author{J. D. Song}
\affiliation{Hefei National Laboratory for Physical Sciences at
Microscale, University of Science and Technology of China, Hefei,
Anhui 230026, People's Republic of China}

\author{S. J. Li}
\affiliation{Hefei National Laboratory for Physical Sciences at
Microscale, University of Science and Technology of China, Hefei,
Anhui 230026, People's Republic of China}

\author{X. F. Sun}
\email{xfsun@ustc.edu.cn}

\affiliation{Hefei National Laboratory for Physical Sciences at
Microscale, University of Science and Technology of China, Hefei,
Anhui 230026, People's Republic of China}

\author{X. G. Li}

\affiliation{Hefei National Laboratory for Physical Sciences at
Microscale, University of Science and Technology of China, Hefei,
Anhui 230026, People's Republic of China}

\affiliation{Department of Physics, University of Science and
Technology of China, Hefei, Anhui 230026, People's Republic of
China}

\date{\today}

\begin{abstract}

We report a study of the Ce doping effect on the thermal conductivity ($\kappa$) of Nd$_{2-x}$Ce$_x$CuO$_4$ (NCCO) at low temperatures down to 0.3 K and in magnetic fields up to 14 T. It is found that with Ce doping, the electronic thermal conductivity increases; at the same time, the $a$-axis field induced changes in $\kappa(H)$, associated with the spin flop and spin polarization of Nd$^{3+}$ sublattice, and the spin flop of Cu$^{2+}$ sublattice, gradually disappear. These are clearly due to the electron doping and the destruction of the antiferromagnetic orders. In the superconducting NCCO with $x$ = 0.14 and 0.18, although the electronic thermal conductivity shows sizable field dependencies with $H \parallel c$, the paramagnetic scattering of phonons is still playing the dominant role in the heat transport, which is different from many other cuprates. In the lightly doped samples ($x$ = 0.03 and 0.06), the low-$T$ $\kappa(H)$ isotherms with $H \parallel c$ show a step-like anomaly and is likely related to the spin/charge stripes.

\end{abstract}

\pacs{74.72.Cj, 66.70.-f, 74.72.Ek, 75.47.-m}

%74.72.Cj Insulating parent compounds
%66.70.-f Nonelectronic thermal conduction and heat-pulse propagation in solids
%74.72.Ek Electron-doped
%75.47.-m Magnetotransport phenomena; materials for magnetotransport

\maketitle

\section{Introduction}

Low-temperature thermal conductivity ($\kappa$) has been extensively studied for high-$T_c$ cuprates, since it provides the most straightforward information of quasiparticle (QP) transport properties and the peculiar electronic state.\cite{Hussey_Review, Graf, Durst, Taillefer, Sutherland_YBCO, Sun_YBCO, Behnia, Chiao, Sun_Bi2212, Takeya, Ando_BSLCO, Hawthorn_Tl, Sun_nonuniversal} In addition, in the underdoped cuprates the heat transport can also show rich physics about the magnon heat transport of low-dimensional spin systems, the phonon heat transport and its interaction with peculiar spin/charge order.\cite{Baberski_LNSCO, Sun_LESCO, Sun_LCO, Sun_PLCCO}

The electronic thermal conductivity in cuprate superconductors has two competitive effects caused by magnetic field: one is the QP scattering by vortices, which is predominant at relatively higher temperatures;\cite{Krishana, Franz, Vekhter, Ando_Bi2212} another one is the field-induced QP excitations, which is known as the ``Volovik" effect and is more important at subKelvin temperatures.\cite{Volovik, Vekhter, Aubin, Chiao_YBCO} In this regard, the $\kappa(H)$ behaviors at subKelvin temperatures, free from the complicated vortice scattering effect, are therefore very useful for probing the nature of ground state.\cite{Sun_YBCO, Ando_BSLCO, Sun_LSCO, Hawthorn_LSCO} One important finding for La$_{2-x}$Sr$_x$CuO$_4$ (LSCO) and Bi$_2$Sr$_{2-x}$La$_x$CuO$_{6+\delta}$ (BSLCO) is that at the lowest temperatures the $\kappa$ is enhanced by magnetic field in the highly doped samples, while it is suppressed by field in the low doped samples.\cite{Sun_LSCO, Ando_BSLCO, Hawthorn_LSCO} This pointed to the metal-to-insulator crossover (MIC) of the ground states of these materials, which were found to be coincided with those revealed by the resistivity measurements under ultra-high magnetic fields.\cite{Sun_LSCO, Ando_BSLCO} Using this correspondence, the MIC of YBa$_2$Cu$_3$O$_y$ (YBCO) was determined by the heat transport measurements.\cite{Sun_YBCO} In these works, the field dependence of $\kappa$ was attributed only to the electron transport and the phononic thermal conductivity was assumed to be independent of magnetic field. However, it was later found that in an undoped compound Pr$_{1.3}$La$_{0.7}$CuO$_4$ (PLCO), the phononic thermal conductivity has rather strong field dependence, which can be understood by paramagnetic scattering of phonons.\cite{Sun_PLCO}

A more interesting case was found in another parent compound of electron doped cuprates, Nd$_2$CuO$_4$ (NCO). It is known that the insulating parent compounds of high-$T_c$ cuprates have an antiferromagnetic (AF) order of Cu$^{2+}$ spins. Because of the quasi-two-dimensionality of the Cu$^{2+}$ spin structure, these materials can show rather strong magnon heat transport at high temperatures.\cite{Sun_LCO, Sun_PLCCO, Hess_LCO, Berggold, Sun_PLCO, Jin_NCO, Zhao_NCO} In NCO, the Cu$^{2+}$ spins order antiferromagnetically below $T_N \sim$ 250 K with a noncollinear magnetic structure.\cite{Matsuda, Endoh, Cherny, Structure, Structure2, Richard, NCO_neutron} When the magnetic field is applied in the CuO$_2$ plane, the Cu$^{2+}$ spins can re-orientate and enter a spin-flop state.\cite{Structure, Structure2, Cherny, Thalmeier, NCO_SF1, NCO_SF2,NCO_SF3, NCO_SF4} The transition fields were reported to be 4.5 and 0.75 T for $H \parallel a$ and $H \parallel [110]$, respectively.\cite{Cherny, NCO_SF3} In addition, because of the strong coupling between Nd$^{3+}$ spins and Cu$^{2+}$ spins, which was found to be about 4 T,\cite{Richard} the Nd$^{3+}$ spins are generally considered to change together with the Cu$^{2+}$ sublattice under the influence of magnetic field. However, at low temperatures ($<$ 1.5 K) when the AF order of Nd$^{3+}$ spins is formed, the magnetic structure of Nd$^{3+}$ sublattice could be changed by the magnetic field independently,\cite{Zhao_NCO} although it may have the same noncollinear spin structure as the Cu$^{2+}$ spins in zero field.\cite{Structure, NCO_neutron, Lynn} It has been found that all these magnetic phase transitions and field-induced transitions of magnetic structures have substantial effects on the low-$T$ heat transport of NCO.\cite{Jin_NCO, Zhao_NCO, Li_NCO}

Based on two earlier studies and our own experiments,\cite{Jin_NCO, Li_NCO, Zhao_NCO} we have shown the main mechanisms of the low-$T$ heat transport of NCO.\cite{Zhao_NCO} In zero field, the low-$T$ thermal conductivity is purely phononic with rather strong scatterings from the paramagnetic moments and the Nd$^{3+}$ magnon excitations. In high magnetic field along either the $c$ axis or the $ab$ plane, the low-$T$ $\kappa$ can be significantly enhanced because of the weakening of magnetic scattering. In addition, the field-induced spin flop or spin polarization of Nd$^{3+}$ sublattice results in drastic changes of $\kappa$ at low fields along the $a$ axis or the [110] direction. With $H \parallel a$ and at subKelvin temperatures, the Nd$^{3+}$ magnons can act as heat carriers in the spin-flopped state, however, their transport can exist only in some intermediate field regime and is suppressed by a succeeding spin-polarization transition for $H \parallel a$. On the basis of these knowledge,\cite{Zhao_NCO} it is called for to study how the heat-transport behavior evolves with increasing the charge carriers by doping Ce into NCO.

In this work, we study the Ce doping effect on the heat transport of Nd$_{2-x}$Ce$_x$CuO$_4$ (NCCO) at low temperatures down to 0.3 K and in magnetic fields up to 14 T. It is found that peculiar behaviors of the non-doped NCO, namely, the $a$-axis field induced changes of $\kappa$, associated with the magnetic transitions of the Nd$^{3+}$ and the Cu$^{2+}$ spin sublattices, gradually disappear with increasing $x$. At the same time, the electron thermal conductivity increases and its field dependencies becomes visible with $H \parallel c$ in the superconducting NCCO ($x$ = 0.14 and 0.18). However, the paramagnetic scattering of phonons is playing the dominant role in the field dependencies of $\kappa$ even in the superconducting samples, which is different from many other cuprates. Another notable finding is that a step-like increase in the $\kappa(H)$ isotherms with $H \parallel c$ for the lightly doped samples ($x$ = 0.03 and 0.06), which may be related to the spin/charge stripes.

\section{Experiments}

High-quality Nd$_{2-x}$Ce$_x$CuO$_4$ ($x =$ 0--0.18) single crystals were grown by using the slow cooling method with CuO$_2$ as a self flux.\cite{Zhao_NCO} The as-grown crystals are plate-like with the $c$ axis along the thickness dimension. For heat transport measurements, the crystals were cut into long-bar shape with the longest dimension along the $a$ axis, by using the x-ray back-reflection Laue photographs. The thermal conductivity was then measured along the $a$ axis by using a ``one heater, two thermometers" technique and two different processes: (i) in a $^3$He refrigerator and a 14 T magnet at temperature region of 0.3 -- 8 K; (ii) in a pulse-tube refrigerator for the zero-field data above 4 K.\cite{Sun_DTN, Wang_HMO, Zhao_GFO, Zhao_NCO, Li_NGSO, Zhang_GdErTO}

The as-grown NCCO crystals usually have excess oxygen, which results in smaller electron concentration than the Ce content ($x$) nominally produces.\cite{Armitage_Review} Therefore, most of the samples, without specially mentioning, were annealed in flowing Ar and at 900 $^\circ$C, accompanied with a slow cooling (2--3 $^\circ$C/min) to the room temperature. This annealing process can effectively remove the excess oxygen, and is necessary for $x \ge$ 0.10 samples exhibiting superconductivity. In this paper, we show the thermal conductivity data of several NCCO single crystals with actual $x$ = 0, 0.03, 0.06, 0.14, and 0.18 (the uncertainties are typically $\pm$0.005). Two methods were used to determine the actual value of $x$. One is a direct composition measurement by using a X-ray fluorescence spectrometry with Rh anode tube (XRF-1800, Shimdzu). Another one is taking the x-ray diffraction data of our samples and making comparison with the relationship between the Ce concentration and the $c$-axis lattice parameter from the literature.\cite{Armitage_Review} The superconductivity of these samples was tested by DC susceptibility measurements using a SQUID magnetometer (Quantum Design). It was found that the $x$ = 0, 0.03, and 0.06 crystals are nonsuperconducting at temperature down to 2 K; the $x$ = 0.14 and 0.18 crystals have $T_c$ = 21 and 12 K, which correspond to the underdoped and overdoped levels, respectively.\cite{Armitage_Review}

\section{Results and Discussion}

\subsection{Temperature dependence of $\kappa$ in zero field}

\begin{figure}
\includegraphics[clip,width=7.5cm]{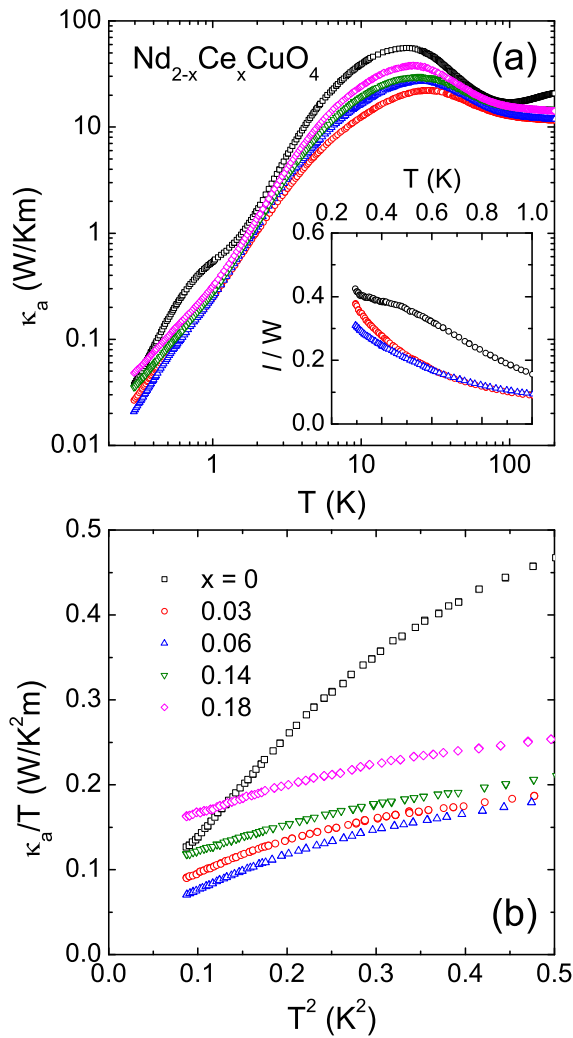}
\caption{(Color online) (a) The $a$-axis thermal conductivity of Nd$_{2-x}$Ce$_x$CuO$_4$ single crystals in zero field in a log-log plot. Inset: the ratios of the phonon mean free path $l$ to the averaged sample width $W$ at $T <$ 1 K for nonsuperconducting samples of $x$ = 0, 0.03, and 0.06. (b) The low-temperature data plotted in $\kappa/T$ vs $T^2$.}
\end{figure}

Figure 1(a) shows the temperature dependencies of $\kappa$ of NCCO single crystals in zero field. The $x$ = 0 sample shows a large phonon peak (56 W/Km) at 20 K,\cite{Zhao_NCO, Berman} of which the magnitude is dominated by the crystal defects/impurities scattering on phonons. This indicates that the quality of our crystals are as good as those grown by using the floating-zone method.\cite{Jin_NCO} Note that a much higher low-$T$ peak of $\kappa(T)$ was observed in a flux-grown NCCO ($x$ = 0.025) crystal.\cite{Cohn_NCCO} Another remarkable feature of the $x$ = 0 data is that an obvious variation in slope of the $\kappa(T)$ curve shows up at around 1.5 K, which is related to the AF ordering of Nd$^{3+}$ ions.\cite{Structure, NCO_neutron, Lynn, Zhao_NCO} It was discussed in our earlier paper that this is caused by an enhanced phonon scattering by the spin fluctuations at the critical region of AF phase transition.\cite{Zhao_NCO} Note that the Nd$^{3+}$ magnon excitations are not able to transport heat at low temperatures ($T <$ 1.5 K), because the spin-anisotropy gap prevents the low-energy magnons from being thermally excited. With doping Ce, the $\kappa(T)$ display several changes. First, the $\kappa$ at high temperatures is strongly suppressed by slight doping of $x$ = 0.03, with the peak value decreasing to 22 W/Km. It could be due to an increase of the impurity scattering on phonons. However, the high-$T$ $\kappa$ is gradually increased with increasing $x$ above 0.03. The low-$T$ $\kappa$ show similar changes although at subKelvin temperatures the magnitude of $\kappa$ has some relationship to not only the microscopic mechanism of heat transport but also the sample size.\cite{Berman, Sun_Comment} One possible reason that the high-$T$ $\kappa$ is enhanced with increasing $x$ is due to the increasing contribution from the electron transport. Second, the concavity at 1.5 K becomes much weaker in the Ce-doped samples, but a small curvature is still observable even in the $x$ = 0.18 sample.

In usual insulators, the microscopic scatterings on phonons are quickly smeared out at low temperatures, and finally a boundary scattering limit can be achieved when the mean free path of phonons is long enough to be the same as the averaged sample width.\cite{Berman} The phononic thermal conductivity can be expressed by the kinetic formula $\kappa_{ph} = \frac{1}{3}Cv_pl$,\cite{Berman} where $C = \beta T^3$ is the phonon specific heat at low temperatures, $\beta$ is a $T$-independent coefficient, $v_p$ is the average velocity and $l$ is the mean free path of phonons. The $\beta$ and $v_p$ values have been known experimentally for NCO,\cite{Specific_heat, Velocity} so the mean free path can be calculated from the $\kappa$.\cite{Sun_Comment, Zhao_NCO} The inset to Fig. 1(a) shows the ratios between $l$ and the averaged sample width $W = 2\sqrt{A/\pi}$,\cite{Zhao_GFO, Zhao_NCO, Berman} where $A$ is the area of cross section, for three insulating samples. It can be seen that the ratios $l/W$ are 0.3--0.4 at 0.3 K, which means that the microscopic scattering on phonons are still not negligible at this temperature region.\cite{Berman, Sun_Comment} Our earlier work on NCO has revealed that the magnetic scattering on phonons is rather strong at low temperatures down to 0.3 K.\cite{Zhao_NCO} Therefore, the calculation of $l$ indicates that the magnetic scattering are comparably significant in these insulating NCCO crystals.

Figure 1(b) shows the low-$T$ data plotted in $\kappa/T$ vs $T^2$. This kind of plot is commonly used for separating the electronic and phononic terms of $\kappa$ in the superconducting samples.\cite{Taillefer, Sutherland_YBCO, Sun_YBCO, Behnia, Chiao, Sun_Bi2212, Takeya, Ando_BSLCO, Hawthorn_Tl, Sun_nonuniversal} The data would show a straight line at very low temperatures if the phonon boundary scattering limit were achieved, which usually requires data at milli-Kelvin temperatures. The slope and the zero-$T$ intercept ($\kappa_0/T$) of the straight line give the phonon conductivity and the electronic thermal conductivity, respectively. As have been well studied for the $p$-type high-$T_c$ cuprates, the non-zero $\kappa_0/T$ indicated the extended low-energy QPs and is a strong evidence for the nodal superconducting gap, like the famous $d_{x^2-y^2}$ symmetry.\cite{Hussey_Review, Graf, Durst} We actually tried the $\kappa$ measurements on NCCO at low temperatures down to 60 mK (as shown in the Appendix). However, for the superconducting samples, the $\kappa/T$ vs $T^2$ plots show strong downward curvature at $T <$ 300 mK (see Fig. 7), which is known to be due to a decoupling between electrons and phonons.\cite{Smith} In other words, it is not feasible to try to get precise electronic term of $\kappa_0/T$ in the electron-doped cuprates. Therefore, the present work focuses on the measurements at $T \ge$ 300 mK, where the electron-phonon decoupling is not serious. From Fig. 1(b), it is qualitatively clear that the QP heat transport is gradually increased with increasing $x$ for the superconducting samples. It is also very likely that the $\kappa/T$ extrapolated to $T =$ 0 K would have non-zero values in the $x$ = 0.14 and 0.18 samples. In this regard, the thermal conductivity data seem to support a nodal gap of NCCO superconductors, similar to that of the $p$-type cuprate superconductors. In passing, it should be noted that the various experimental investigations have not yet arrived at a consistent conclusion on the symmetry of the superconducting gap in the electron-doped cuprates.\cite{Armitage_Review}

\subsection{Thermal conductivity of lightly doped NCCO with $x =$ 0.03 and 0.06}

\begin{figure}
\includegraphics[clip,width=7.0cm]{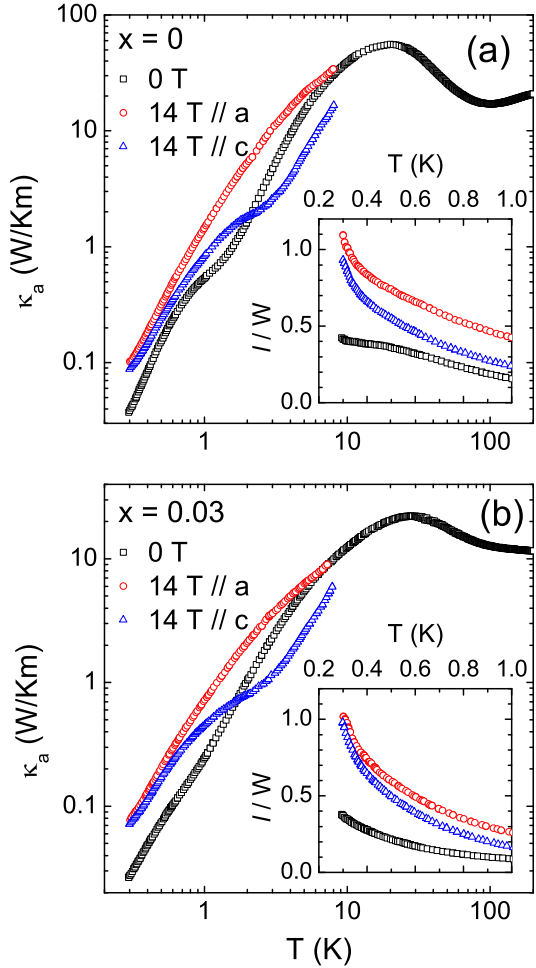}
\caption{(Color online) Comparison of $\kappa(T)$ between the $x$ = 0 and 0.03 Nd$_{2-x}$Ce$_x$CuO$_4$ single crystals with the zero field and 14 T field applied along the $a$ or the $c$ axis. The insets show the ratio of the phonon mean free path $l$ to the averaged sample width $W$ at $T <$ 1 K and in different fields.}
\end{figure}

The effect of slight doping ($x$ = 0.03) on the $\kappa(T)$ is demonstrated in Fig. 2. As mentioned above, the zero-field $\kappa(T)$ of the $x$ = 0 sample shows a concavity at about 1.5 K, which is due to the Nd$^{3+}$ AF transition. This transition is completely suppressed in 14 T along the $a$ axis, which is strong enough to polarize the Nd$^{3+}$ spins. In this case, as shown in the inset to Fig. 2(a), the phonon mean free path is much larger than that in the zero field and approaches the averaged sample width at 0.3 K. It indicates that the boundary scattering limit of phonons is nearly achieved in the 14 T field ($\parallel a$) and at such low temperatures.\cite{Zhao_NCO, Zhao_GFO, Sun_Comment} Namely, there is strong magnetic scattering of phonons in zero field and the scattering is substantially reduced in high field.\cite{Zhao_NCO} When a 14 T field is applied along the $c$ axis, the $\kappa$ displays a complex dependence on temperature. First, the $\kappa$ in the 14 T $c$-axis field are always smaller than those in the 14 T $a$-axis field. Second, another even stronger concavity shows up in the $\kappa(T)$ curve at higher temperatures about 3 K. It should have different origin from the zero-field one since it appears at much higher temperatures than the N\'eel temperature of Nd$^{3+}$ spins. Resonant scattering of phonons by magnetic excitations likely remains active. The details have been carefully discussed in our earlier work.\cite{Zhao_NCO}

In the $x$ = 0.03 sample, the concavity related to the Nd$^{3+}$ AF transition becomes much weaker and shifts to lower temperatures. In zero field, it is located at $T <$ 1 K, suggesting that the N\'eel transition of Nd$^{3+}$ spins is somewhat suppressed by the 3 \% replacement of Nd$^{3+}$ with Ce$^{4+}$ ions. The 14 T fields along the $a$ or $c$ axis bring similar changes of $\kappa$ to those of the $x$ = 0 sample. Note that there is also a strong concavity at about 3 K when the 14 T field is along the $c$ axis. Moreover, compared with the doping effect on the N\'eel transition of Nd$^{3+}$ spins, this 3 K concavity is robust against the slight Ce doping. It thus indicates that the 3 K concavity is related to a paramagnetic scattering of phonons, which can be hardly modified by the 3 \% replacement of Nd$^{3+}$ with Ce$^{4+}$.

\begin{figure}
\includegraphics[clip,width=8.5cm]{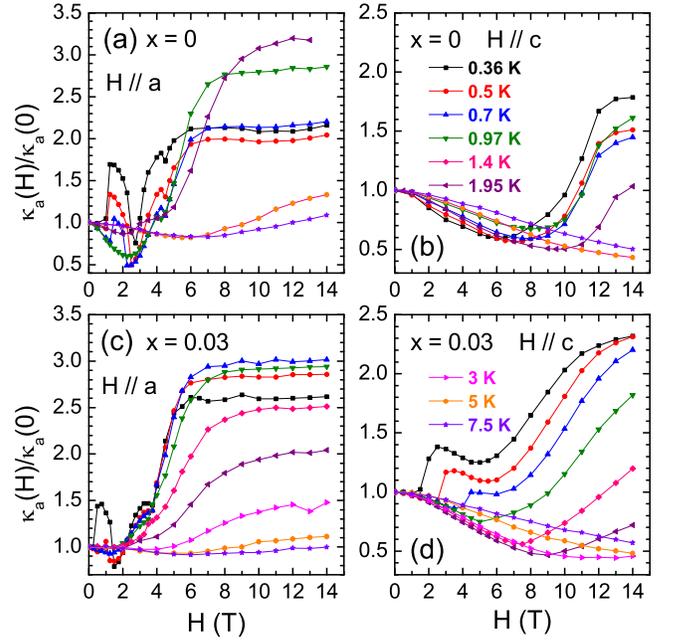}
\caption{(Color online) Comparison of the magnetic-field dependencies of thermal conductivity between the $x$ = 0 and 0.03 Nd$_{2-x}$Ce$_x$CuO$_4$ single crystals with the field applied along the $a$ or the $c$ axis.}
\end{figure}

The effect of slight doping ($x$ = 0.03) on the $\kappa$ is further demonstrated in Fig. 3 with the detailed magnetic-field dependencies. It is known that the $\kappa(H)$ of the $x$ = 0 sample has two mechanisms.\cite{Zhao_NCO} The first one is the paramagnetic scattering effect on phonons and is present in both the in-plane field and the $c$-axis field.\cite{Sun_PLCO, Sun_GBCO, Li_NGSO} This scattering causes a broad valley-like behavior of $\kappa(H)$, as Fig. 3(b) mainly shows. The second one is related to the magnetic transitions of the Nd$^{3+}$ and Cu$^{2+}$ spins induced by the in-plane field. As shown in Fig. 3(a), with increasing field along the $a$ axis, a step-like increase and a rather sharp dip observed at $\sim$ 1 and 2.5 T in subKelvin $\kappa(H)$ isotherms are due to the spin-flop and spin-polarization transitions of Nd$^{3+}$ sublattice, respectively; another small dip at 4.5 T is due to the spin flop of Cu$^{2+}$ sublattice.\cite{Zhao_NCO} It can be seen from Fig. 3(c) that in the case of the $a$-axis field, the slight Ce doping changes the $\kappa(H)$ behaviors not so drastically. Mainly, all the three transition fields become smaller and are about 0.5, 1.5 and 3.75 T at 0.36 K. Apparently, the Ce doping disturbs both the Nd$^{3+}$ and Cu$^{2+}$ spin sublattices. In contrast, the slight Ce doping changes the $\kappa(H)$ with $H \parallel c$ more seriously. As shown in Fig. 3(d), the 0.36-K $\kappa(H)$ exhibits a sharp increase at 1.5 T and a broad valley at 5 T, and both of them shift to higher fields with increasing temperature. It seems that the latter one is the same as the $\kappa(H)$ behavior of the $x$ = 0 sample in the $c$-axis fields, of which the 0.36 K data show a broad valley at 6 T. Therefore, the low-field increase of $\kappa$ is a new feature induced by the slight Ce doping. The origin will be discussed in the following text.

Here we provide a brief description of the paramagnetic scattering effect discussed previously for NCO.\cite{Zhao_NCO} It has been known for a long time that the paramagentic ions can effectively scatter phonons when the Zeeman splitting of the lowest spin states is comparable to the phonon energy.\cite{Berman} Although it is difficult to model this process and give quantitative results with the paramagnetic scattering, a simplified calculation could show a qualitative behavior of $\kappa(H)$.\cite{Sun_GBCO} That is, if the lowest spin states are not split in zero field, the $\kappa$ first decreases and then increases with increasing field, and finally recovers its zero-field value  at high-field limit.\cite{Sun_GBCO} Furthermore, the position of the $\kappa(H)$ minimum shifts to higher field with increasing temperature.\cite{Sun_GBCO} This phenomenon has been observed in many magnetic oxides.\cite{Sun_PLCO, Sun_GBCO, Li_NGSO} In NCO, the ``free" spins at very low temperatures can be either the spin vacancies/defects on the long-range-ordered Cu$^{2+}$ spin lattice or those on the ordered Nd$^{3+}$ spin lattice. Since the low-$T$ $\kappa$ in the high-field limit is apparently larger than those in zero field, it is likely that the ``free" spins on the Nd$^{3+}$ sites, whose ground-state doublet can be split in the zero field,\cite{Structure} rather than the Cu$^{2+}$ free spins,\cite{Sun_PLCO} are responsible for scattering phonons.

\begin{figure}
\includegraphics[clip,width=8.5cm]{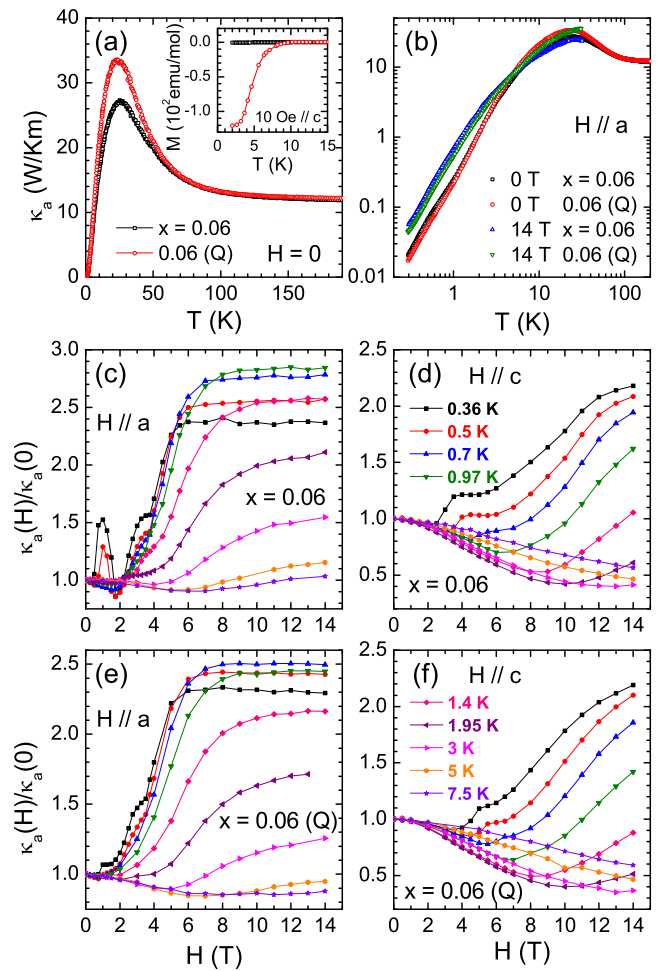}
\caption{(Color online) Thermal conductivity of two $x$ = 0.06 Nd$_{2-x}$Ce$_x$CuO$_4$ single crystals, which were annealed with different processes. The one labelled by ``$x$ = 0.06" was annealed in the same way as other NCCO samples, while the other one labelled by ``$x$ = 0.06 (Q)" was annealed in flowing Ar and at 900 $^\circ$C but quenched to the liquid nitrogen temperature. (a) The $\kappa(T)$ in linear plot. Inset: the magnetic susceptibilities measured in 10 Oe along the $c$ axis after cooling the samples in zero field. The data show that the $x$ = 0.06 sample is not superconducting above 2 K and the $x$ = 0.06 (Q) sample has $T_c =$ 9 K. (b) The log-log plot of $\kappa(T)$ in both zero and 14 T fields along the $a$ axis. (c-f) Low-temperature $\kappa(H)$ isotherms for two samples in magnetic fields along either the $a$ or the $c$ axis.}
\end{figure}

Figure 4 shows the $\kappa(T)$ and $\kappa(H)$ data of two $x$ = 0.06 Nd$_{2-x}$Ce$_x$CuO$_4$ single crystals, which were annealed with different processes. The magnetic susceptibility data in the inset of Fig. 4(a) indicates that the $x$ = 0.06 sample, annealed in flowing Ar and at 900 $^\circ$C, is nonsuperconducting, which is consistent with the known phase diagram of NCCO.\cite{Armitage_Review} The $x$ = 0.06 (Q) sample, annealed under the same condition but followed with a quenching to the liquid nitrogen temperature, has superconducting transition at $T_c =$ 9 K, which indicates a smaller oxygen content and a higher electron concentration in this quenched sample. In this regard, the larger $\kappa$ of the $x$ = 0.06 (Q) sample at high temperatures, as shown in Fig. 4(a), is consistent with the evolution of $\kappa$ with increasing $x$ shown in Fig. 1(a). In these 6 \% Ce doped NCCO, the changes of the $\kappa(T)$ slopes at about 1 K are even weaker than that in the $x$ = 0.03 sample. Applying 14 T along the $a$ axis can also significantly enhance the low-$T$ $\kappa$, which is similar to the case of $x$ = 0.03 sample.

Figures 4(c-f) show the detailed field dependencies of $\kappa$ in two $x =$ 0.06 samples, which behave rather similarly to the $x$ = 0.03 sample. For $H \parallel a$, the anomalies of $\kappa$ associated with the Nd$^{3+}$ spin flop, the Nd$^{3+}$ spin polarization, and the Cu$^{2+}$ spin flop occur at comparable fields in the two differently doped NCCO. The main difference between these two $x$ dopings is that the low-field increase of $\kappa$ in the case of $H \parallel c$ is less sharp and weaker, and occurs at higher field in the $x$ = 0.06 samples. This demonstrates that although this step-like transition is caused by Ce doping, it however becomes weaker with increasing $x$. For the $x$ = 0.06 (Q) sample, which has higher electron concentration, all the anomalies in the $\kappa(H)$ isotherms display further evolution. In the case of $H \parallel a$, the Nd$^{3+}$ spin-flop induced increase of $\kappa$ and the Nd$^{3+}$ spin-polarization related minimum almost disappears, and the kink caused by Cu$^{2+}$ spin flop, however, changes only slightly. Furthermore, the step-like increase of $\kappa$ in the case of $H \parallel c$ becomes even weaker and shifts to higher field. In this regard, getting higher electron doping in the quenched $x$ = 0.06 sample plays a similar role in the heat transport to doping more Ce.

It should be noted that the kink induced by the Cu$^{2+}$ spin flop is still present in the low-$T$ $\kappa(H)$ of the superconducting $x$ = 0.06 (Q) sample, with a small shift to lower field. This indicates that the long-range AF order of Cu$^{2+}$ spins is not destroyed in this sample. It is a new experimental evidence supporting the coexistence of AF order and superconductivity in the underdoped $n$-type cuprates.\cite{{Armitage_Review}}

\subsection{Thermal conductivity of highly doped NCCO with $x =$ 0.14 and 0.18}

\begin{figure}
\includegraphics[clip,width=8.5cm]{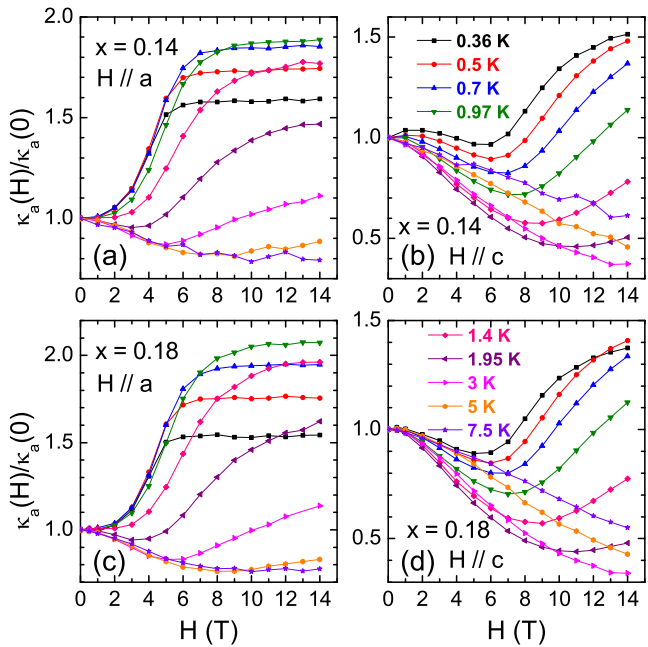}
\caption{(Color online) Magnetic-field dependencies of thermal
conductivity for two superconducting Nd$_{2-x}$Ce$_x$CuO$_4$ single crystals with $x$ = 0.14 and 0.18.}
\end{figure}

Figure 5 shows the $\kappa(H)$ isotherms of superconducting NCCO single crystals with $x$ = 0.14 and 0.18. Compared with the lower Ce dopings, these two samples show simpler field dependencies of $\kappa$, particularly in the low field regions, for both $H \parallel a$ and $H \parallel c$. Actually, all the anomalies in the $\kappa(H)$ curves for $H \parallel a$, which are related to the magnetic transitions of Nd$^{3+}$ or Cu$^{2+}$ sublattices, completely disappear. It is easy to understand since these Ce dopings are high enough to destroy the long-range AF orders of both the Nd$^{3+}$ and Cu$^{2+}$ sublattices.\cite{Armitage_Review} The low-field anomaly for $H \parallel c$ is also completely suppressed. However, the field dependencies of $\kappa$ are still quite strong in these two samples. Then, what is the main mechanism for the heat-transport behaviors of these superconducting samples?

First, as shown in Figs. 5(a) and 5(c), the $\kappa$ with $H \parallel a$ of these two samples have comparably strong field dependencies, which is not coincided with the QP transport properties of high-$T_c$ cuprates. It is well known that all the high-$T_c$ cuprates, including NCCO, have strong anisotropy of the upper critical field, that is, $H_{c2}$ for $H \parallel a$ is much larger than that for $H \parallel c$. For NCCO, the $H_{c2}$ is smaller than 10 T for $H \parallel c$ and is about 10 times larger for $H \parallel a$ at very low temperatures.\cite{Hidaka, Wang} This means that the magnetic fields up to 14 T are still far below the $H_{c2}$ for $H \parallel a$ and the possible QP contribution to $\kappa$ in Figs. 5(a) and 5(c) should be much smaller than that to $\kappa$ with $H \parallel c$, for which 14 T is strong enough to completely suppress the superconductivity. It is also notable that at very low temperatures the $\kappa$ in $H \parallel a$ shows high-field plateau, which is actually a common phenomenon for all the NCCO crystals. This behavior in NCO was discussed to be mainly due to the suppression of phonon-paramagnetic scattering.\cite{Zhao_NCO} Apparently, this feature is common to all NCCO crystals.

Second, the field dependencies of $\kappa$ in $H \parallel c$ also behave rather differently to those in other high-$T_c$ cuprates. The very-low-$T$ $\kappa(H)$ have been well studied for LSCO, BSLCO, and YBCO systems with hole concentrations varying from underdoped to overdoped regimes,\cite{Sun_LSCO, Hawthorn_LSCO, Ando_BSLCO, Sun_YBCO} and the field-induced changes of $\kappa$ are attributed to the QP transport properties. A common result is that at the lowest temperatures the $\kappa$ is enhanced by magnetic field in the highly doped samples, while it is suppressed by field in the low doped samples. This pointed to the MIC of the ground states of these materials.\cite{Sun_LSCO, Ando_BSLCO, Sun_YBCO} In those earlier studies on the $p$-type cuprates, the magnetic field were always applied along the $c$ axis, that is, perpendicular to the CuO$_2$ planes, in which direction the upper critical field is much lower. The NCCO $\kappa(H)$ data with $H \parallel c$ are, however, very different from the $p$-type cuprate superconductors. As shown in Figs. 5(b) and 5(d), upon increasing field at very low temperatures, the $\kappa$ firstly decrease; after going through a broad minimum, the $\kappa$ gradually increase in high fields and the 14 T data are clearly larger than the zero-field $\kappa$. In fact, these $\kappa(H)$ behaviors are very similar to the parent compound with $x$ = 0 (see Fig. 3(b)). As already discussed, this result of NCO can be well explained by the paramagnetic scattering effect on phonons.\cite{Zhao_NCO, Sun_PLCO, Sun_GBCO, Li_NGSO} Therefore, it seems that the field dependencies of the electron term of $\kappa$ are playing a minor role in the $\kappa(H)$ of superconducting NCCO, although the QP heat transport is naturally expected to be affected by magnetic fields.

Third, although the paramagnetic scattering is likely playing a dominant role in the $\kappa(H)$ behaviors of the superconducting NCCO, the significant difference of $\kappa(H)$ between $H \parallel a$ and $H \parallel c$ is not understandable in this mechanism. As already demonstrated in many other AF materials,\cite{Sun_PLCO, Sun_GBCO, Li_NGSO} the paramagnetic scattering would result in a qualitatively isotropic behavior for field along different directions. It is therefore due to a non-negligible field dependence of electronic thermal conductivity in the case of $H \parallel c$. It is rather clear that the electronic thermal conductivity with $H \parallel c$ is a decreasing function of $H$ in the superconducting NCCO, which is similar to the underdoped $p$-type cuprates.\cite{Sun_LSCO, Hawthorn_LSCO, Ando_BSLCO, Sun_YBCO} In passing, it should be emphasized that in NCO the paramagnetic scattering effect is dominant with $H \parallel c$, which has negligible effect on the AF states of Nd$^{3+}$ and Cu$^{2+}$ sublattices; whereas, in the superconducting NCCO the paramagnetic scattering is more evidenced with $H \parallel a$, which has weak effect on the superconducting state (the AF orders are destroyed and irrelevant). Furthermore, the $\kappa(H)$ behaviors caused by the paramagnetic scattering in these two cases look rather different. It can be due to the modifications in the crystal-field levels of Nd$^{3+}$ ions, induced by the Ce doping.

\subsection{Comparison of differently doped NCCO}

\begin{figure}
\includegraphics[clip,width=7.0cm]{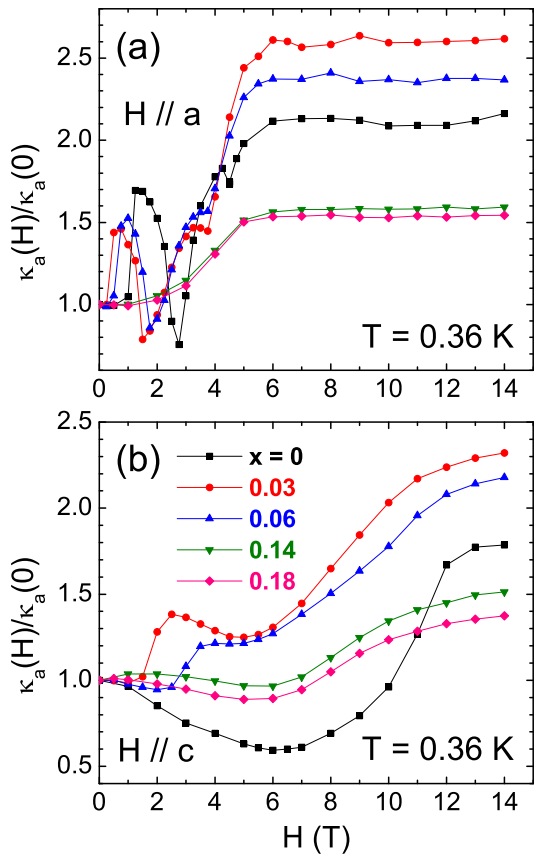}
\caption{(Color online) Comparison of the lowest-$T$ $\kappa(H)$ data among differently doped Nd$_{2-x}$Ce$_x$CuO$_4$ single crystals.}
\end{figure}

Figure 6 shows a comparison of the $\kappa(H)$ data at 0.36 K among all the NCCO crystals we studied. Based on the above discussions, it is known that the electron doping induced changes of $\kappa$ are visible mainly in the case of $H \parallel c$, while the data with $H \parallel a$ are caused by the magnetic contributions. First, in the case of $H \parallel a$, the highly-doped samples show a gradual increase of $\kappa$ and a high-field plateau with increasing field, which is due to a weakness of the paramagnetic scattering. In the non-doped and lightly-doped samples, besides this change trend, there are clearly several low-field anomalies caused by the magnetic transitions of Nd$^{3+}$ and Cu$^{2+}$ spins, which have long-range AF orders. Second, in the case of $H \parallel c$, the data for non-doped sample should be solely explained by the paramagnetic scattering, while the data for highly-doped superconducting samples must include some changes of the electronic term. However, it is by now difficult to separate the field-induced changes of the electronic thermal conductivity from the raw data, due to the dominant paramagnetic scattering effect on phonons. Third, a remarkable result is the step-like increase of $\kappa$ in the lightly-doped samples ($x$ = 0.03 and 0.06) with $H \parallel c$. Note that this phenomenon should have no direct relationship to the dopants, Ce$^{4+}$ ions, since they are nonmagnetic. It is therefore clear that this anomaly must be related to some kind of the electronic behavior, and all other mechanisms such as paramagnetic scattering and field-induce magnetic transitions are not relevant. One possibility is that the electron self-organized structures could be playing a role.\cite{Kivelson}

It is well known that the $p$-type cuprates have static spin/charge stripes in the low doping region.\cite{Kivelson, Tranquada, Yamada, Matsuda_stripes, Mook, Ando_stripes, Valla, Tranquada2} The stripe phase is actually a periodic distribution of antiferromagnetically-ordered spin regions
separated by quasi-one-dimensional charged domain walls, which act
as magnetic antiphase boundaries. Since the nonuniform charge distribution
is expected to induce variations of the local crystal structure,\cite{Tranquada, Mook} which disturb phonons, the phonon heat transport is expected to be capable of detecting stripes. It was firstly evidenced in rare-earth and Sr co-doped La$_2$CuO$_4$, such as La$_{1.28}$Nd$_{0.6}$Sr$_{0.12}$CuO$_4$ and La$_{1.88-y}$Eu$_y$Sr$_{0.12}$CuO$_4$, that the dynamical stripes cause a pronounced damping of phonon heat transport, while the well ordered static stripes do not suppress the phonon transport so significantly.\cite{Baberski_LNSCO, Sun_LESCO} Then, it was demonstrated that the $c$-axis phonon heat transport is a good probe of the stripe formation in the lightly hole-doped LSCO; namely, the stripes in this system are well-ordered in the CuO$_2$ planes but are disordered along the $c$ axis,\cite{Matsuda} which causes the $c$-axis phonons to be strongly scattered.\cite{Sun_LCO} The stripe scenario was also proposed for the $n$-type cuprates, but has not been proved by enough experimental results.\cite{Sun_PLCCO, Zamborszky_PCCO, Fournier_PCCO, Dai_PLCCO} An indirect evidence is the data of anisotropic thermal conductivity in Pr$_{1.3-x}$La$_{0.7}$Ce$_x$CuO$_4$ system.\cite{Sun_PLCCO} It was found that the $c$-axis phonon heat transport was quickly suppressed by slight Ce doping, which is the same as the Sr-doping effect in LSCO,\cite{Sun_LCO} and was understood as better ordering of the stripes in the CuO$_2$ planes than along the $c$ axis. In the present case, the $c$-axis heat transport of NCCO cannot be studied for these thin-plate shaped samples. However, the very-low-$T$ $\kappa(H)$ data seem to be able to give some signature of the spin/charge stripes. It is very likely that the stripes (static) in the zero field are not ordered very well and cause some scattering on phonons; applying magnetic field can drive them to form a more ordered state\cite{Tranquada2, LeBoeuf} and therefore remove the scattering effect. In this regard, the direct comparison between the heat transport and other experiments like neutron scattering, however, is not yet available for NCCO and called for further experimental investigations.

At last, we need to point out that the magnetic-field dependencies of $\kappa$ of NCCO are probably the most special case in high-$T_c$ cuprates. In the $p$-type cuprates of LSCO, BSLCO, Bi$_2$Sr$_2$CaCu$_2$O$_{8+\delta}$ (Bi2212), and YBCO, the field dependencies of $\kappa$ seem to be mainly due to the QP transport properties, and there is by now no clear evidence for whether the phonon transport can be strongly affected by magnetic field. Apparently, these materials have no magnetic ions except for Cu$^{2+}$. On the one hand, the AF order of Cu$^{2+}$ spins is completely suppressed in their superconducting phase. On the other hand, the possible paramagnetic scattering, which can be related to Cu$^{2+}$ free spins produced by defects or impurities, cannot be significant if the samples are high-quality single crystals.\cite{Sun_PLCO, Moler, Revaz, Wright, Chen, Brugger, Schottky} The impact of the latter one can be referred to the result of PLCO, a parent compound of $n$-type cuprates, in which the magnetic scattering caused by the Cu$^{2+}$ free spins could change the phononic thermal conductivity about 20 \%.\cite{Sun_PLCO} In the present case of NCCO, however, the magnetic scattering has a much larger effect on the field dependencies of $\kappa$, since this system has much more magnetic ions.

\section{Conclusions}

In summary, we study the heat transport of electron-doped cuprates Nd$_{2-x}$Ce$_x$CuO$_4$ at low temperatures down to 0.3 K and in magnetic fields up to 14 T. It has been known that in the parent material Nd$_2$CuO$_4$, the low-$T$ thermal conductivity is purely phononic with rather strong scatterings from the paramagnetic moments and the Nd$^{3+}$ magnons. With Ce doping, the low-field changes of $\kappa$ with $H \parallel a$, associated with the spin flop and spin polarization of Nd$^{3+}$ sublattice, and the spin flop of Cu$^{2+}$ sublattice, gradually disappear; at the same time, the electronic thermal conductivity increases. However, even in the superconducting NCCO with highly doped Ce ($x$ = 0.14 and 0.18), the magnetic scattering of phonons is playing the dominant role in the heat transport, which is different from many other cuprates, although the field dependence of electronic thermal conductivity is distinguishable in the data with $H \parallel c$. Another interesting finding is a step-like increase of $\kappa$ ($H \parallel c$) in the lightly doped samples ($x$ = 0.03 and 0.06), which may be related to the spin/charge stripes.

\begin{acknowledgements}

This work was supported by the National Natural Science Foundation of China, the National Basic Research Program of China (Grants No. 2011CBA00111 and 2012CB922003), and the Fundamental Research Funds for the Central Universities (Programs No. WK2340000035 and WK2030220014).

\end{acknowledgements}

\section*{Appendix: Measurements of $\kappa$ in milli-Kelvin temperature regime}

\begin{figure}
\includegraphics[clip,width=7.5cm]{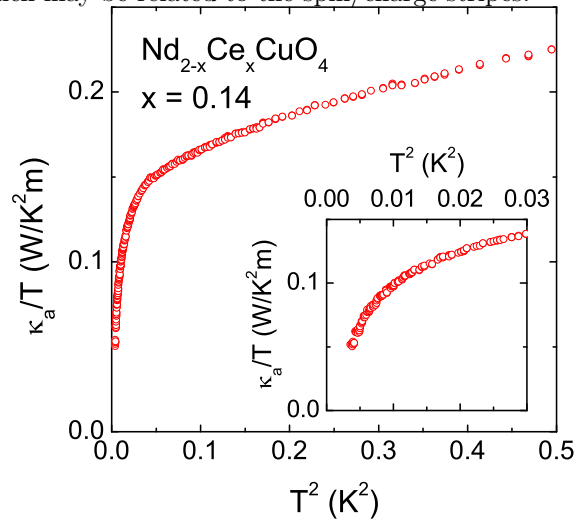}
\caption{(Color online) The $a$-axis thermal conductivity of Nd$_{2-x}$Ce$_x$CuO$_4$ ($x =$ 0.14) single crystal in the zero field, plotted in $\kappa/T$ vs $T^2$. The size of this sample is 2.00 $\times$ 0.83 $\times$ 0.069 mm$^3$. Inset: The zoom-in of the plot at very low temperatures.}
\end{figure}

We have tried to measure the thermal conductivity of superconducting NCCO single crystals at ultra-low temperatures ($<$ 100 mK) using a ``one heater, two thermometers" technique in a dilution refrigerator.\cite{Sun_YBCO, Sun_nonuniversal, Zhou_ZCO} Figure 7 shows a representative result for $x$ = 0.14, which is superconducting at 21 K. The most remarkable feature of the data is a strong downturn at low temperatures ($<$ 200 mK for this sample). As a result, the $\kappa/T$ is heading to a zero value at $T \rightarrow$ 0, as the inset to Fig. 7 show. However, this is known as an extrinsic phenomenon of the QP heat transport and was discussed to be caused by an electron-phonon decoupling.\cite{Smith}

\end{document}